# Development and Deployment of Fixed Wireless Access in South West Nigeria: Performance and Evaluation

[1]Oluwaranti Adeniran and [2]Achimugu Philip

**Abstract:** Fixed Wireless Access (FWA) involves the use of wireless technology to replace copper to connect subscribers to the telephone network. It is a variant of wireless broadband which provides an alternative in the so-called 'last mile' connectivity between the subscriber and the fixed telecommunications network. FWA could either be narrowband or broadband and it is predominantly deployed using the Code Division Multiple Access (CDMA) technology. In assessing the extent of development and deployment of FWA, the perspective of the operators and users was elicited primarily through the use of questionnaires. Issues like setup cost, tax, Government incentive, availability of infrastructure and manpower applied to the operators while on the users' part factors like quality of service, signal strength as well as call rate were considered. The South western zone of Nigeria is regarded as one of the most urbanized regions in the south of Sahara, this is not out of place considering the fact that Lagos which is the nation's commercial nerve centre and a mega-city is located within this region. The scope of this research covered this very lively part of the country. The relationship between the parameters or variables considered was established using an appropriate statistical method: The Regression analysis. In terms of users' preference, Global System of Mobile communication (GSM) was compared with FWA. Results were interpreted and suitable conclusions were drawn to wrap up a quite revealing work.

**Index Terms:** Fixed Wireless Access (FWA), Global Systems for Mobile communications (GSM), Code Division Multiple Access (CDMA), Quality of Service (QoS), telecommunication networks

———————————— ◆ ————————————

## 1 Introduction

Access to Information and effective means of communication are fundamental issues in every human society. In African societies that existed before colonial rule, people communicated using various instruments and codes such as talking drums, flutes, gongs, town crier and village square meetings [1]. Many historical accounts are still preserved on the walls of caves and especially through oral tradition. The use of writing and the invention of printing transformed the type and content of recorded history. Communications on a universal scale became possible through the use of books, newspapers, magazines and radio. More recent technological innovations have increased further the reach and speed of communications, culminating for now with digital technology [3].

The rapid and unprecedented growth in telecommunications and information technology elsewhere in the world is now reaching saturation point and attention is shifting to Africa as one of the last major markets for telecommunications. In Nigeria, international businessmen and businesswomen, various arms of government, educational institutions and private individuals are demanding high quality and easily available telecommunications services to the same standards as are available internationally. With a population of over 120 million, Nigeria is home to one out of every five people that inhabit sub-Saharan Africa and as such represents 20% of the telecommunications market in that area [5]. To this end, the Government now promotes the development of Fixed Wireless Access throughout the country. The growth of wireless broadband networks is expected to gradually outpace landline communications as advancements in these technologies are enabling higher broadband speeds [1].

Fixed Wireless Access removes the need to drape wires across the country or dig up roads to provide fixed telecommunication links, as is the case for fixed telephony and cable networks. It would provide fast, always on access to the Internet, high capacity data transfer, on-line banking cum shopping and many other services. As a result, it can easily also provide an effective platform from which to expand existing infrastructure, or serve to provide infrastructure in hitherto under-served areas.

————————————————

1. *Oluwaranti Adeniran is with the Department of Computer Science and Engineering at Obafemi Awolowo University, Ile-Ife, Nigeria.*

2. *Achimugu Philip is with the Department of Computer Science of Lead City University, Ibadan, Nigeria.*

Simply put, Fixed Wireless Access (FWA) is the use of wireless technology to replace copper to connect subscribers to the telephone network [8] as shown in Fig. 1.

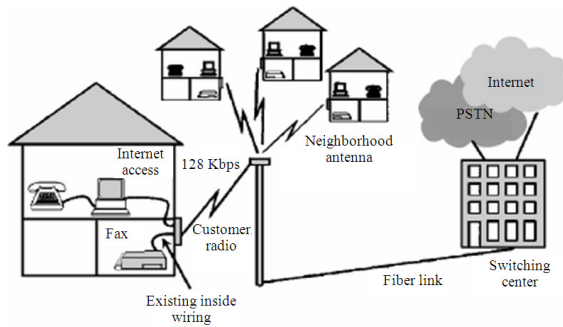

Fig. 1: Fixed wireless access configuration. Source: Muller (2003)

### 1.1 Fixed wireless access in Nigeria:

The structure of the fixed line sector has changed in step with the changes in political administration. According to [2], Nigerian Telecommunications Limited (NITEL), the state-owned fixed line operator, had an installed capacity of some 720,270 lines in year 2000 but only around half of these (497,070) were connected. There has been little or no roll out since then, with a penetration rate of some 0.43%. Official statistics are subject to due diligence, but perhaps half of these connected lines are still analogue. This dilapidated network is highly unreliable. The sector was first deregulated in 1992 by the Nigerian Communications Commission decree, which facilitated the entry of the private sector. In order to meet the demand for communications, nearly thirty licenses were issued under the military regime of General Sani Abacha to Private Telecommunications Operators (called PTOs). The classes of licenses granted under the decree to these 'PTOs' included Private Network Links (PNL), community telephony, public payphones and Value Added Network Services (VANS).

Not all of these have become operational, but those which have deployed a variety of Wireless Local Loop (WLL) networks using Advanced Mobile Phone System/Service (AMPS), Time Division Multiple Access (TDMA), Global System for Mobile communications (GSM) and Code Division Multiple Access (CDMA) technologies are mostly located in Lagos (the commercial capital).

With the passing of the military regime in 1999, government policy called for the issuance of four GSM licenses, the privatization of NITEL, the awarding of a second carrier licenses and the issuance of FWA licenses. Each of these licenses came with high roll-out obligations to achieve much higher tele-density. Because the government was unable to invest the heavy capital expenditure required, it has sought to create an environment conducive to the private sector. The NCC has succeeded in turning the telecommunications sector around. It has rationalized the ad-hoc licensing regime and revoked inactive licenses; it is tackling the once haphazard allocation of radio spectrum by imposing structure onto the sector. The regulator has provided a much greater degree of certainty for both foreign and Nigerian investors.

## 2. Materials and Methods

### 2.1 Questionnaire design/distribution:

Open and closed response categories exist in questionnaires design. Two types of questionnaires were used for this study: User questionnaire and Operator questionnaire. The questionnaire for the users was divided into two parts: Personal information and appraisal of service. The personal information part provided information on each respondent that uses a fixed wireless access; five questions were asked in this part which covered respondent's background and monthly income. The appraisal of service section which asked 14 questions provided information about each respondent's perception of the quality of service being provided by the service providers.

The questionnaire for the service providers was designed to elicit information on the factors that affect the provision of fixed wireless service (voice, data, Multimedia Message Service, Internet access). This questionnaire was also divided into three main sections: Company Profile, Respondent's Data and Operational Data. Company's Profile section provided information on the company's name and year of establishment. The respondent's data section provided information about the status of the respondent in the company, while the operational data section provided respondent-based information about Government policy, operating environment and the availability of technical support.

Table 1: Summary of questionnaires distributed and returned

| State | Total No of questionnaires | No of questionnaires returned | Percentage returned |
|---|---|---|---|
| Ekiti | 30 | 30 | 100.00 |
| Lagos | 400 | 283 | 70.75 |
| Ogun | 70 | 58 | 82.80 |
| Ondo | 100 | 100 | 100.00 |
| Osun | 100 | 88 | 88.00 |
| Oyo | 300 | 252 | 83.33 |
| Total | 1000 | 811 | 81.10 |

The population distribution is summarized in Table 1. Samples were taken from Lagos and the capital cities of the other five States. The response rate for the user questionnaire was 80% while that of operator was 98%.

## 3. Results

### 3.1 Operator's response:

According to [7] there are 24 FWA operators in Nigeria. However, this questionnaire focused on the major players in the FWA industry. To this end, a total of 5 companies were sampled. Mathematically, this accounts for approximately 20% of the total operators. However, this percentage concurs with Pareto's principle of which the abridged version states that: "20% of the operators provide 80% of the services." In evaluating their performances in terms of development (growth), questions like:

How many states in the South West have been covered by specific operators? Government policy on the FWA sector (tax, licensing fee and so on); Setup cost incurred by FWA operators. Level of challenge(s) faced that could impede development as well as the effect of the Global System for Mobile telecommunications (GSM) on FWA competition were considered and answered accordingly in this sub section.

In addition, for the area of deployment, the time taken to roll out coverage would be the index for appraisal.

The five operators considered were: Starcomms, Multilinks, Intercellular, MTS First and Odua Tel. It is therefore interesting to note that ALL the six states in the South-Western region of Nigeria have been covered by these operators. Specifically, Odua Tel and Multilinks have covered five states each while MTS First has covered three states. Starcomms has two states and finally, Intercellular has only one state covered. Thus an average of three states has been covered by each operator considered; therefore, growth has reached three states on the average.

### 3.2 Government policies:

Herein entails government policies on operators. Policies such as licensing fees, tax, incentives and legislations were taken into consideration. Table 2 shows the response. From the table 2, about 43% of the operator claimed that the licensing fee is high. On the issue of tax, it was a tie on medium and high. Government incentives to operators are minimal as over 85% of the respondents claimed. Legislation is appropriate as can be seen from the table 2.

### 3.3 Cost of setup:

Setup cost for the organization is one of the most important factors that dictate the development of FWA. The analysis of this factor would serve as a forerunner to the regression analysis consisting of setup cost and government policies.

Further analysis would now be carried out on the relationship between Setup cost and the various government policies listed above. The regression analysis would suffice for this.

From Table 3, the regression value, R, is 0.69. A good analysis has R tending towards 1 which shows a strong relationship between the dependent and independent variables. From table 3 (coefficient table), the regression model equation is:

$$Y = 0.6363 + 0.2500_{licencing} + 0.4091_{Tax} - 0.7272_{govt.incentives} \quad (1),$$

where, Y = Dependent variable: Setup cost.

### 3.4 User's response:

Without a demand for goods and services, supply would be needless. Thus, for FWA operators to be running in the black, their services must be met with demands.

The population size (N) for the users was 750. This would used for the analysis. In evaluating the performance of fixed wireless access in the South West region, the following questions would serve as guide:

- What are the age bracket of respondents and their level of education? This would help to ascertain if the questionnaires were given to the right set of people
- What is the income bracket of respondents? This question would go hand in hand with the cost of acquisition of the service and the tariffs charged by operators. With this, the affordability of the services deployed by the operators would be ascertained

Table 2: Response on licensing fee, tax, government incentives and legislation respectively

|  |  | Frequency | Percent | Valid percent | Cumulative percent |
|---|---|---|---|---|---|
| **Licensing fee** |  |  |  |  |  |
| Valid | EXHORBITANT | 2 | 28.6 | 33.3 | 33.3 |
|  | HIGH | 3 | 42.9 | 50.0 | 83.3 |
|  | MEDIUM | 1 | 14.3 | 16.7 | 100.0 |
|  | Total | 6 | 85.7 | 100.0 |  |
| Missing | System | 1 | 14.3 |  |  |
| Total |  | 7 | 100.0 |  |  |
| **Tax** |  |  |  |  |  |
| Valid | EXHORBITANT | 1 | 14.3 | 20.0 | 20.0 |
|  | HIGH | 2 | 28.6 | 40.0 | 60.0 |
|  | MEDIUM | 2 | 28.6 | 40.0 | 100.0 |
|  | Total | 5 | 71.4 | 100.0 |  |
| Missing | System | 2 | 28.6 |  |  |
| Total |  | 7 | 100.0 |  |  |
| **Government incentives** |  |  |  |  |  |
| Valid | MINIMAL | 6 | 85.7 | 100.0 | 100.0 |
| Missing | System | 1 | 14.3 |  |  |
| Total |  | 7 | 100.0 |  |  |
| **Existing Legislation** |  |  |  |  |  |
| Valid | APPROPRIATE | 3 | 42.9 | 75.0 | 75.0 |
|  | INADEQUATE | 1 | 14.3 | 25.0 | 100.0 |
|  | Total | 4 | 57.1 | 100.0 |  |
| Missing | System | 3 | 42.9 |  |  |
| Total |  | 7 | 100.0 |  |  |

Table 3: Regression table for relationship between setup cost and government incentives, licensing fee and tax
(a) Model summary[b]

| Model | R | $R^2$ | Adjusted $R^2$ | Std. error of the estimate |
|---|---|---|---|---|
| 1 | 0.693[a] | 0.481 | -0.039 | 0.917 |

[a]: Predictors: (Constant), government incentives, licencing fee, tax; [b]: Dependent variable: Cost of setup

(b) Coefficients[a]

|  | Unstandardized coefficients | Standardized coefficients |
|---|---|---|

| Model | | B | Std. error | Beta | t | Sig. |
|---|---|---|---|---|---|---|
| 1 | (Constant) | 0.63636 | 0.997 | | 0.638 | 0.569 |
| | Licensing fee | 0.25000 | 0.794 | 0.192 | 0.315 | 0.774 |
| | Tax | 0.40909 | 0.732 | 0.344 | 0.559 | 0.615 |
| | Government incentives | -0.72727 | 0.782 | -0.394 | 0.930 | 0.421 |

a: Dependent variable: Cost of setup

- How good or bad are the service signal and the quality of service? This helps to give a good appraisal of the services deployed by the operators

  Finally, the level of acceptability of FWA as compared to the GSM

### 3.5 Age bracket of respondent:

Table 4 shows the percentage age distribution of respondents to the questionnaires. It also shows the level of education. Over 67% of the users were between the ages of 18 and 40. This gives a good working sample for the analysis as this is the age bracket of the working class of growing any nation. More so, over 68% of the population is educated to the tertiary level. With these assertions, a thorough analysis can now be done with the other questions listed above.

### 3.6 Income level vs. cost of service acquisition and tariff charged:

As stated earlier, this analysis would help in evaluating how the income level of users affects the usage of the FWA services rolled out by operators. Ideally, it is expected that the lower the income, the less patronage of services (even essential ones like communication). Going by this belief, the tariff charges were found to be moderate since subscribers on the average are comfortable. Therefore, it is clear that FWA is highly affordable in purchase and use.

### 3.7 Service signal and quality of service appraisal:

Here, evaluation of how strong or weak the service signal of providers would be discussed. Also to be discussed is the quality of network service. The pie charts in Fig. 2 give a pictorial form of these. From the first chart, 49.2% of the users claim that the service signal is good

Table 4: Frequency distribution of respondent's age and education level

| | | Frequency | Percent | Valid percent | Cumulative percent |
|---|---|---|---|---|---|
| **Age bracket** | | | | | |
| Valid | Below 18 | 90 | 12.0 | 12.1 | 12.1 |
| | 18-40 | 506 | 67.5 | 67.8 | 79.9 |
| | Over 40 | 150 | 20.0 | 20.1 | 100.0 |
| | Total | 746 | 99.5 | 100.0 | |
| Missing | system | 4 | 0.5 | | |
| Total | | 750 | 100.0 | | |

**Highest level of education**

| | | | | | |
|---|---|---|---|---|---|
| Valid | No formal education | 5 | 0.7 | 0.7 | 0.7 |
| | Primary | 16 | 2.1 | 2.2 | 2.9 |
| | Secondary | 195 | 26.0 | 26.7 | 29.6 |
| | Tertiary | 513 | 68.4 | 70.4 | 100.0 |
| | Total | 729 | 97.2 | 100.0 | |
| Missing | system | 21 | 2.8 | | |
| Total | | 750 | 100.0 | | |

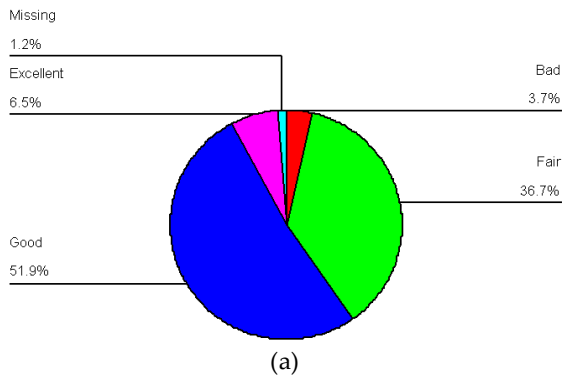

(a)

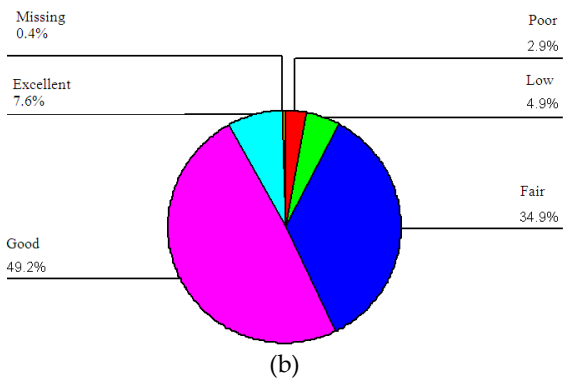

(b)

Fig. 2: Pie-charts for signal strengths and quality of service (a) quality of service (b) signal strength

and over 51% gave a good response on the quality of service. The major deduction from this is: Operators have done well in providing good FWA to their customers though there is room for improvement. That is, in spite of the setbacks discussed in the operator's section; service provision is still worthwhile.

### 3.8 Level of acceptability of FWA against GSM:
This would be concluded by looking at the level of acceptability of FWA in the region. Approximately 40% of the users preferred fixed wireless as the technology for voice calls. The margin between these users and the others that prefer GSM is about 16% (56% preferred the GSM). The major reason for this difference is simply the technology behind the two networks. GSM has far more coverage than the FWA. The right word is mobility. The GSM allows users 'roaming' capabilities Fig. 3.

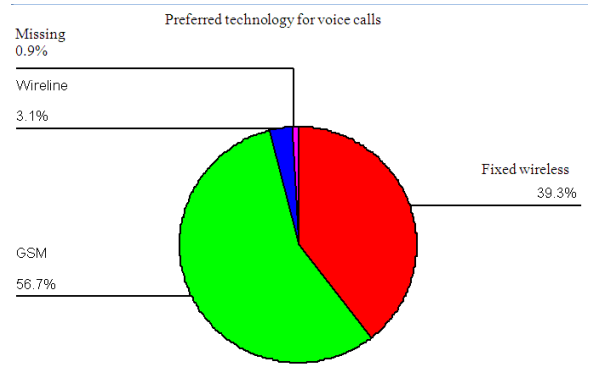

Fig. 3: Level of acceptability of AWA against GSM

## 4 Discussion
The deployment of fixed wireless access technology in the south western zone of Nigeria has been quite commendable in spite of the very difficult terrain which the Nigerian environment offers. Government policy and incentive has generally been less than desirable. Infrastructural decadence has contributed immensely to the cost of deployment and also the rate charged for services. The Set up cost for fixed wireless access is regarded to be exorbitant by most of the operators. It also became glaring from this research that two of the factors which contribute to rapid deployment of FWA: availability of Infrastructure and manpower are readily available while the same does not apply to the availability of technology.

FWA cannot be said to be a reserve of the rich in the six states covered as the use of the technology on the general note was not affected by the income bracket of the respondents. In accessing the affordability of FWA, the income level, cost of acquisition and tariffs charged are the parameters considered. Most of the respondents regard FWA as highly affordable and a sizeable percentage agrees that it is a cheaper alternative to communication (in terms of call rate) relative to the GSM technology. However, GSM is the preferred technology for voice calls due to the level of mobility offered by its roaming capability.

The quality of service offered by the operators was evaluated in terms of Signal strength of networks and quality of service of the networks. The effort of the Operators was rated as good though a few users believe they have performed poorly.

## 5. Conclusion

Operators of FWA are definitely not finding things convenient as the various analyses have shown. From taxes paid to problems of power supply, security of installations and so on, it is obvious that development and deployment would be held back. Deployment is directly related to development as without growth, there can be no services to roll out. Meanwhile, subscribers are quite comfortable with the signal strength and quality of service fixed wireless access.

**Dr. Adeniran Oluwaranti** is a Senior Lecturer at the Obafemi Awolowo University, Ile-Ife, Nigeria. He can be reached at Computer Building, Room 117; Department of Computer Science and Engineering, Faculty of Technology, Obafemi Awolowo University, Ile-Ife, Nigeria. He holds B.Sc, M.Sc and PhD degrees in Computer Science from the Obafemi Awolowo University, Ile-Ife. His research interest includes wireless networks and communication. He has published in many reputable journals and refereed conferences both within and outside Nigeria.

**ACHIMUGU Philip** is a Lecturer at the Department of Computer Science of Lead City University, Ibadan, Nigeria. He holds B.Sc and M.Sc degrees in 2004 and 2009 respectively in Computer Science. He is also a PhD student at the Obafemi Awolowo University, Ile-Ife, Nigeria in the same field. He has over 18 publications in reputable journals and refereed learned conferences both home and abroad. His research area is mainly software engineering with emphasis on: Development techniques, Development tools, Software products architecture and Usability.